\documentclass[useAMS,usenatbib]{mn2e}
\usepackage{graphicx}
\newcommand{\pr}{\ensuremath{P_{\rm r}}}

\title[Asteroid polarimetric properties]
{A polarimetric study of asteroids:\\ Fitting phase - polarization curves
\thanks{Partly based on observations carried out 
at the Complejo Astron\'omico El Leoncito, operated under agreement between the Consejo 
Nacional de Investigaciones Cient\'{\i}ficas y T\'ecnicas de la Rep\'ublica Argentina 
and the National Universities of La Plata, C\'ordoba, and San Juan.}}
\author[A. Cellino et al.]{A.\ Cellino$^{1}$, S.\ Bagnulo$^{2}$, 
R.\ Gil-Hutton$^{3}$, P.\ Tanga$^{4}$, M.\ Ca\~nada-Assandri$^{3}$, 
\newauthor
and E. F. Tedesco$^{5}$\\
$^{1}$INAF - Osservatorio Astrofisico di Torino, I-10025 Pino Torinese, Italy. 
{\rm E-mail: cellino@oato.inaf.it}\\
$^{2}$Armagh Observatory, College Hill, Armagh BT61 9DG, UK. {\rm E-mail: sba@arm.ac.uk}\\
$^{3}$CASLEO and San Juan National University, San Juan, Argentina. 
{\rm E-mail: rgilhutton@casleo.gov.ar, micanada03@casleo.gov.ar}\\
$^{4}$Observatoire de la C\^ote d'Azur, Nice, France. {\rm E-mail: Paolo.Tanga@oca.eu}\\
$^{5}$Planetary Science Institute, Tucson, USA. {\rm E-mail: eft@psi.edu}
}

\begin{document}


\pagerange{\pageref{firstpage}--\pageref{lastpage}} \pubyear{2015}

\maketitle

\label{firstpage}

\begin{abstract}
By considering all published asteroid linear polarization data available
in the literature, it is possible to obtain updated phase - polarization
curves for several tens of objects. In a separate paper (Cellino et al., 2015a, 
MNRAS, 451, 3473) we have produced new calibrations of different relations between 
the geometric albedo and several polarimetric parameters, based on an analysis of a 
limited sample of asteroids for which the albedo is known with sufficient
accuracy. In this paper, we present the main polarization parameters and 
corresponding albedos for a larger dataset of asteroids which we did not use for 
calibration purposes. We find a good agreement between the albedo values computed
using different polarization parameters. Conversely, in the case of the so-called Barbarian
asteroids the situation is rather unclear. Moreover, we present an updated analysis
of the distributions of different polarimetric parameters, including the so-called 
inversion angle and the solar phase angle corresponding to the extreme value of negative 
polarization, and study their mutual relations. We find that the above parameters
can be used to clearly distinguish some unusual classes of asteroids. Polarimetric 
parameters are known to be related to physical properties of asteroid surfaces which 
are difficult to infer by means of other observing techniques. By using a much larger
dataset, in our analysis we confirm and extend some results obtained in the past by other 
authors, and we explore more systematically some features that had been mostly 
unexplored before, mainly concerning the morphology of the negative polarization branch. 
\end{abstract}

\begin{keywords}
polarization -- minor planets, asteroids: general.
\end{keywords}

\section{Introduction}\label{intro}
In a separate paper \citep[][hereinafter Paper I]{paperI}, we have performed an extensive
analysis of the problem of finding satisfactory calibrations of several
relations between different polarimetric parameters and the geometric albedo.
Our analysis was based on using for calibration purposes a sample
of asteroids for which we have reliable independent estimates of the albedo,
based on accurate measurements of their size, reliable estimates of their
absolute magnitude, and using the known relation linking the size, geometric albedo and 
absolute magnitude of an asteroid, namely:
\begin{equation}
\log(D) = 3.1236 -0.2H - 0.5 \log(p_V)
\end{equation} \label{Eqn:DHpv}
where $D$ is the diameter expressed in km (assuming the object
is spherical), $H$ is the absolute magnitude (in the Johnson $V$ band by definiton) 
and $p_V$ is the geometric albedo (again, in the Johnson $V$ band). 
A list of asteroids suited for the purposes of calibration was published
by Shevchenko and Tedesco (2006), hereinafter \citet{ShevTed}. 
In recent years we have performed campaigns of polarimetric
observations of objects belonging to the \citet{ShevTed} list, in order
to obtain for them accurate polarimetric measurements. Paper I presented the 
results of our analysis of these asteroids. 
By analyzing the phase -polarization curves of the objects, namely the variation of the
linear polarization $P_r$ as a function of the solar phase angle\footnote{$P_r$ is the degree of 
linear polarization with a sign that is defined to be positive when the plane of 
polarization is found to be perpendicular to the Sun - observer - target plane 
(scattering plane), and negative when the plane of polarization is parallel to the 
scattering plane. The solar phase angle, herein after referred to simply as the ``phase angle'', 
is the angle between the directions to the Sun and to the observer, as seen from the
target object.}, we computed
a variety of polarimetric parameters that are known to be diagnostic
of albedo, including the polarimetric slope $h$ and the extreme value of
negative polarization $P_{\rm min}$ (for an introduction to the basic notions
of asteroid polarimetry, that we will not repeat here, see Paper I). In addition,
we also considered some new polarimetric parameters, including the $p\ast$ parameter
suggested by \citet{Masiero12}, as well as a new parameter, 
named $\Psi$, that we introduced in Paper I. 

In order to derive accurate values of the above polarimetric parameters from
available data, in Paper I we focused our analysis on asteroids for which 
we have a good number of polarimetric measurements satisfactorily sampling the
phase - polarization curves, and we made use of 
the following exponential-linear relation to fit the phase - polarization
curves of the objects under scrutiny:
\begin{equation}
\pr = A (e^{-\alpha/B} - 1) + C \cdot \alpha \label{Eqn:ABC}
\end{equation}
where $\alpha$ is the phase angle expressed in degrees, and $A$, $B$,
$C$ are parameters whose values have to be determined by means of best-fit
techniques. The above analytical representation has been found in the past
to be well suited to fit phase - polarization curves \citep{Muinonenetal09}.
Note that this relation does not take into account the possible presence 
of a polarization surge at very small phase angles. This effect, found 
by \citet{Rosenbush05} and \citet{Rosenbush09} to
be possibly present in the case of a couple of very-high albedo asteroids,
(64) Angelina and (44) Nysa, respectively, is rather
negligible for the purposes of the present analysis. We do not analyze Angelina
because we already did it in Paper I. We note also that the available 
measurements obtained for this object at very small phase angles  
have error bars too large to be accepted by our severe selection criteria adopted
in Paper I.
In the case of Nysa, which is analyzed in the present paper, the few measurements 
suggesting a surge of negative
polarization at phase angles $<2^\circ$ do not seem to produce any important
consequence on the overall fit of its phase - polarization curve, although
we will see below that this asteroid seems to be rather peculiar in some aspects.

Using the exponential-linear relation, we were able to find suitable calibrations of 
several relations between the geometric albedo and polarimetric parameters, generally 
described in the form:
\begin{displaymath}
\log(p_V) = A \log(w) + B
\end{displaymath}
where $p_V$ is the geometric albedo in $V$ light and $w$ is one of
several possible polarimetric parameters characterizing the
morphology of available phase - polarization curves. 

In Paper I, that was exclusively focused on the issue of the calibration of different
albedo - polarization relations, we analyzed $22$ 
asteroids belonging to the \citet{ShevTed} list. However,
we could also compute the most relevant polarimetric parameters for a larger number
of other asteroids, not included in the above-mentioned list, because they were not 
suitable for the specific purposes of Paper I. In this paper, we present our 
results for 
the remaining objects. In particular, we consider all available polarimetric data,
taken from different sources, including the PDS\footnote{Data
  available at http://pds.jpl.nasa.gov/ (files maintained by
  D.F. Lupishko and I.N. Belskaya)}, and recent papers
\citep{RGH11, RGH12, Assandrietal2012, Giletal14}.  
For each object we computed different estimates of the albedos using the same methods
used in Paper I. 

Moreover, in the present paper we also present a more extensive analysis of the 
distribution of different polarimetric parameters among the asteroid population,
using all available data, and we
also analyze some interesting relations between different polarimetric parameters. 
This was beyond the scope of the analysis
performed in Paper I, but we investigate now such relations, including some that
in the past were considered to be directly diagnostic of surface
properties, including the typical sizes of regolith particles.

\section{Polarization parameters and geometric albedos}\label{albedo}
In Paper I (Tables 2, 3, and 5) we listed a summary of several polarimetric 
parameters and corresponding albedo values obtained for $22$ objects from  
\citet{ShevTed}. These asteroids, for which we made an effort to 
obtain new polarimetric meaurements, were chosen for the purposes of calibration 
of different possible albedo - polarization relations. 
Here we consider an additional set of $64$ asteroids that are not
included in \citet{ShevTed}. These asteroids were chosen from those that, in our 
judgment, have phase - polarization data of sufficiently good
quality to derive the major polarimetric parameters with an accuracy high enough
for use in the present study. In terms
of requirements concerning data quality and coverage of the phase - polarization curves, 
we made our selection using the same criteria already adopted and described in Paper I.

Table 1 shows for these objects the polarimetric slope $h$, resulting from the 
computation of a simple linear fit of a minimum of five $P_r$ measurements obtained 
at phase angles larger than $14^\circ$, as well as some other parameters considered 
in Paper I. Apart from $h$, these parameters were obtained from fitting an 
exponential - linear fit (Eq.$~$\ref{Eqn:ABC}) to the whole phase - polarization 
curves. These parameters include another independent estimate of the polarimetric slope,
that we called $h_{ABC}$, obtained as the first derivative of the exponential -
linear curve computed at the inversion angle $\alpha_{inv}$;
the extreme value of negative polarization $P_{\rm min}$; the $\Psi$ parameter 
introduced in Paper I (defined as the difference between the values of $P_r$ 
formally corresponding to phase angles of $30^\circ$ and $10^\circ$, respectively, according
to the best-fit of Eq.$~$\ref{Eqn:ABC}), and the $p\ast$ parameter defined 
by \citet{Masiero12}. The low associated uncertainty of the inversion angle $\alpha_{inv}$
comes also from the best-fit of the exponential-linear relation described in Paper I.  

\begin{table*} 
\label{Tab:Tab1}
\caption{Summary of the formal solutions for the polarimetric parameters for all asteroids 
not included in the Shevchenko and Tedesco (2006) list, for which
we have a suitable coverage of the phase - polarization curves. Each asteroid is identified 
by its number. The second column gives the number $N_{obs}$ of polarimetric measurements used 
in the analysis. 
For the meaning of the other parameters, see the text. The same Table for asteroids included in the
\citet{ShevTed} list have been published in Paper I.} 
\tiny
\begin{tabular}{rrccccccc}
      &                   &                &                &                  &                  &                   & \\
\multicolumn{1}{c}{Number} & \multicolumn{1}{c}{$N_{obs}$} & \multicolumn{1}{c}{$h$} & \multicolumn{1}{c}{$\alpha_{inv}$} & 
\multicolumn{1}{c}{$\alpha(P_{\rm min})$} &
\multicolumn{1}{c}{$P_{\rm min}$} & \multicolumn{1}{c}{$\psi$} & \multicolumn{1}{c}{$h_{ABC}$} & \multicolumn{1}{c}{$P\ast$} \\
      &                   & $\%/^\circ$  & $(^\circ)$ & $(^\circ)$ & \% & \% & $\%/^\circ$ & \% \\ 
      &                   &                &                &                  &                  &                   & \\
    5 & 18 & 0.0953 $\pm$ 0.0038 & 19.91 $\pm$ 0.02 &  7.95 $\pm$ 0.09 & -0.71 $\pm$ 0.02 &  1.758 $\pm$ 0.017 & 0.0953 $\pm$ 
		0.0009 & -0.897 $\pm$ 0.026 \\
    6 & 17 & 0.0945 $\pm$ 0.0047 & 22.01 $\pm$ 0.02 &  8.68 $\pm$ 0.08 & -0.81 $\pm$ 0.01 &  1.641 $\pm$ 0.023 & 0.0968 $\pm$ 
		0.0014 & -0.865 $\pm$ 0.027 \\
    7 & 19 & 0.1099 $\pm$ 0.0028 & 21.35 $\pm$ 0.02 &  8.97 $\pm$ 0.05 & -0.75 $\pm$ 0.01 &  1.761 $\pm$ 0.010 & 0.1024 $\pm$ 
		0.0005 & -0.833 $\pm$ 0.022 \\
    9 & 18 & 0.0680 $\pm$ 0.0063 & 23.01 $\pm$ 0.02 &  8.36 $\pm$ 0.14 & -0.71 $\pm$ 0.01 &  1.210 $\pm$ 0.040 & 0.0712 $\pm$ 
		0.0026 & -1.014 $\pm$ 0.040 \\
   10 & 13 & \multicolumn{1}{c}{--} & 18.21 $\pm$ 0.02 &  7.50 $\pm$ 0.08 & -1.49 $\pm$ 0.02 &  4.568 $\pm$ 0.053 & 0.2298 $\pm$ 
	0.0020 & \multicolumn{1}{c}{--} \\
   11 & 8 & 0.1194 $\pm$ 0.0038 & 18.91 $\pm$ 0.02 &  7.78 $\pm$ 0.10 & -0.75 $\pm$ 0.03 &  2.158 $\pm$ 0.020 & 0.1119 $\pm$ 
	0.0010 & -0.804 $\pm$ 0.024 \\
   12 & 17 & 0.1303 $\pm$ 0.0020 & 23.01 $\pm$ 0.02 & 11.08 $\pm$ 0.07 & -0.80 $\pm$ 0.01 & 1.294 $\pm$ 0.010 & 0.1294 $\pm$ 
	0.0005 & -0.758 $\pm$ 0.019 \\
   13 & 9 & 0.2649 $\pm$ 0.0060 & 21.59 $\pm$ 0.02 &  9.24 $\pm$ 0.12 & -2.06 $\pm$ 0.03 &  4.840 $\pm$ 0.069 & 0.2863 $\pm$ 
	0.0046 & -0.264 $\pm$ 0.017 \\
   14 & 7 & 0.0405 $\pm$ 0.0035 & 14.21 $\pm$ 2.27 & \multicolumn{1}{c}{--} & \multicolumn{1}{c}{--} & \multicolumn{1}{c}{--} &
	\multicolumn{1}{c}{--} & \multicolumn{1}{c}{--} \\
   15 & 6 & \multicolumn{1}{c}{--} & 20.65 $\pm$ 0.02 &  8.86 $\pm$ 0.46 & -0.71 $\pm$ 0.02 &  1.841 $\pm$ 0.144 & 0.1033 $\pm$ 
	0.0075 & \multicolumn{1}{c}{--} \\
   16 & 25 & 0.1207 $\pm$ 0.0041 & 22.45 $\pm$ 0.02 &  8.46 $\pm$ 0.03 & -1.02 $\pm$ 0.01 &  1.887 $\pm$ 0.010 & 0.1109 $\pm$ 
	0.0005 & -0.720 $\pm$ 0.022 \\
   18 & 17 & 0.1052 $\pm$ 0.0015 & 21.69 $\pm$ 0.02 &  8.82 $\pm$ 0.06 & -0.82 $\pm$ 0.01 &  1.762 $\pm$ 0.007 & 0.1035 $\pm$ 
	0.0005 & -0.827 $\pm$ 0.020 \\
   19 & 15 & 0.2458 $\pm$ 0.0045 & 21.49 $\pm$ 0.02 & 10.06 $\pm$ 0.03 & -1.75 $\pm$ 0.01 & 4.843 $\pm$ 0.024 & 0.2860 $\pm$ 
	0.0013 & -0.334 $\pm$ 0.016 \\
   20 & 13 & 0.0771 $\pm$ 0.0093 & 19.45 $\pm$ 0.02 &  6.09 $\pm$ 0.12 & -0.69 $\pm$ 0.02 &  1.312 $\pm$ 0.020 & 0.0681 $\pm$ 
	0.0010 & -0.978 $\pm$ 0.048 \\
   21 & 21 & 0.1275 $\pm$ 0.0022 & 25.07 $\pm$ 0.02 &  9.67 $\pm$ 0.06 & -1.27 $\pm$ 0.01 &  1.936 $\pm$ 0.011 & 0.1285 $\pm$ 
	0.0010 & -0.642 $\pm$ 0.019 \\
   24 & 11 & \multicolumn{1}{c}{--} & 20.45 $\pm$ 0.02 & 9.62 $\pm$ 0.06 & -1.43 $\pm$ 0.02 & 4.555 $\pm$ 0.087 & 0.2475 $\pm$ 
	0.0043 & \multicolumn{1}{c}{--} \\
   29 & 6 & \multicolumn{1}{c}{--} & 21.91 $\pm$ 0.02 &  9.12 $\pm$ 0.12 & -0.87 $\pm$ 0.01 &  1.903 $\pm$ 0.041 & 0.1137 $\pm$ 
	0.0027 & \multicolumn{1}{c}{--} \\
   30 & 8 & 0.0948 $\pm$ 0.0028 & 20.63 $\pm$ 0.02 &  8.11 $\pm$ 0.16 & -0.79 $\pm$ 0.02 &  1.785 $\pm$ 0.017 & 0.0994 $\pm$ 
	0.0012 & -0.872 $\pm$ 0.024 \\
   40 & 18 & 0.0874 $\pm$ 0.0075 & 20.91 $\pm$ 0.02 &  7.15 $\pm$ 0.24 & -0.83 $\pm$ 0.03 &  1.565 $\pm$ 0.048 & 0.0849 $\pm$ 
	0.0031 & -0.886 $\pm$ 0.038 \\
   42 & 9 & 0.1161 $\pm$ 0.0057 & 17.67 $\pm$ 0.02 &  6.85 $\pm$ 0.12 & -0.70 $\pm$ 0.03 &  2.023 $\pm$ 0.027 & 0.1018 $\pm$ 
	0.0013 & -0.832 $\pm$ 0.029 \\
   43 & 5 & 0.0746 $\pm$ 0.0026 & 19.10 $\pm$ 0.99 & \multicolumn{1}{c}{--} & \multicolumn{1}{c}{--} & \multicolumn{1}{c}{--} & 
	\multicolumn{1}{c}{--} & \multicolumn{1}{c}{--} \\
   44 & 26 & 0.0278 $\pm$ 0.0039 & 17.83 $\pm$ 0.02 & 2.86 $\pm$ 0.31 & -0.34 $\pm$ 0.01 & 0.490 $\pm$ 0.013 & 0.0245 $\pm$ 
	0.0006 & -1.513 $\pm$ 0.060 \\
   54 & 7 & \multicolumn{1}{c}{--} & 21.51 $\pm$ 0.02 &  9.44 $\pm$ 0.08 & -1.98 $\pm$ 0.01 &  4.873 $\pm$ 0.046 & 0.2880 $\pm$ 
	0.0027 & \multicolumn{1}{c}{--} \\
   56 & 9 & 0.2728 $\pm$ 0.0050 & 19.39 $\pm$ 0.02 &  9.10 $\pm$ 0.03 & -1.46 $\pm$ 0.01 &  5.321 $\pm$ 0.021 & 0.2665 $\pm$ 
	0.0008 & -0.345 $\pm$ 0.014 \\
   63 & 6 & \multicolumn{1}{c}{--} & 18.67 $\pm$ 0.02 &  6.10 $\pm$ 0.56 & -0.85 $\pm$ 0.10 &  1.778 $\pm$ 0.074 & 0.0919 $\pm$ 
	0.0039 & \multicolumn{1}{c}{--} \\
   71 & 5 & \multicolumn{1}{c}{--} & 16.25 $\pm$ 0.02 & 3.56 $\pm$ 0.73 & -0.66 $\pm$ 0.07 & 1.173 $\pm$ 0.016 & 0.0587 $\pm$ 
	0.0008 & \multicolumn{1}{c}{--} \\
   77 & 5 & \multicolumn{1}{c}{--} & 23.41 $\pm$ 0.02 & 7.22 $\pm$ 1.03 & -1.16 $\pm$ 0.11 & 1.710 $\pm$ 0.164 & 0.0937 $\pm$ 
	0.0148 & \multicolumn{1}{c}{--} \\
   84 & 5 & \multicolumn{1}{c}{--} & 20.87 $\pm$ 0.02 &  9.77 $\pm$ 0.05 & -1.52 $\pm$ 0.01 &  4.565 $\pm$ 0.020 & 0.2567 $\pm$ 
	0.0010 & \multicolumn{1}{c}{--} \\
   87 & 9 & \multicolumn{1}{c}{--} & 21.21 $\pm$ 0.02 & 10.11 $\pm$ 0.06 & -1.02 $\pm$ 0.01 &  3.031 $\pm$ 0.064 & 0.1744 $\pm$ 
	0.0033 & \multicolumn{1}{c}{--} \\
   92 & 5 & \multicolumn{1}{c}{--} & 23.63 $\pm$ 0.02 &  7.31 $\pm$ 0.10 & -0.78 $\pm$ 0.02 &  1.146 $\pm$ 0.013 & 0.0631 $\pm$ 
	0.0006 & \multicolumn{1}{c}{--} \\
   97 & 7 & 0.1676 $\pm$ 0.0056 & 22.19 $\pm$ 0.02 & 10.50 $\pm$ 0.04 & -1.23 $\pm$ 0.01 & 3.244 $\pm$ 0.021 & 0.2026 $\pm$ 
	0.0012 & -0.553 $\pm$ 0.020 \\
  113 & 7 & 0.0762 $\pm$ 0.0071 & 18.45 $\pm$ 0.02 &  7.22 $\pm$ 0.11 & -0.52 $\pm$ 0.03 &  1.429 $\pm$ 0.027 & 0.0736 $\pm$ 
	0.0016 & -1.055 $\pm$ 0.045 \\
  115 & 6 & \multicolumn{1}{c}{--} & 21.09 $\pm$ 0.02 &  8.76 $\pm$ 0.14 & -0.70 $\pm$ 0.01 &  1.646 $\pm$ 0.021 & 0.0944 $\pm$ 
	0.0013 & \multicolumn{1}{c}{--} \\
  128 & 9 & \multicolumn{1}{c}{--} & 19.93 $\pm$ 0.02 &  9.34 $\pm$ 0.33 & -1.50 $\pm$ 0.02 &  5.067 $\pm$ 0.191 & 0.2649 $\pm$ 
	0.0071 & \multicolumn{1}{c}{--} \\
  131 & 7 & \multicolumn{1}{c}{--} & 20.83 $\pm$ 0.02 & 10.09 $\pm$ 0.14 & -0.46 $\pm$ 0.01 & 1.483 $\pm$ 0.036 & 0.0821 $\pm$ 
	0.0017 & \multicolumn{1}{c}{--} \\
  132 & 10 & 0.1328 $\pm$ 0.0041 & 19.09 $\pm$ 0.02 &  6.60 $\pm$ 0.12 & -1.14 $\pm$ 0.05 &  2.464 $\pm$ 0.035 & 0.1290 $\pm$ 
	0.0016 & -0.659 $\pm$ 0.025 \\
  138 & 8 & 0.1063 $\pm$ 0.0114 & 18.97 $\pm$ 3.01 & \multicolumn{1}{c}{--} & \multicolumn{1}{c}{--} & \multicolumn{1}{c}{--} & 
	\multicolumn{1}{c}{--} & \multicolumn{1}{c}{--} \\
  161 & 5 & \multicolumn{1}{c}{--} & 19.43 $\pm$ 0.02 &  7.60 $\pm$ 0.03 & -1.25 $\pm$ 0.05 &  3.135 $\pm$ 0.039 & 0.1669 $\pm$ 
	0.0022 & \multicolumn{1}{c}{--} \\
  172 & 10 & 0.1173 $\pm$ 0.0042 & 28.11 $\pm$ 0.02 & 12.82 $\pm$ 0.24 & -1.43 $\pm$ 0.01 &  1.703 $\pm$ 0.019 & 0.1705 $\pm$ 
	0.0072 & -0.641 $\pm$ 0.023 \\
  186 & 5 & \multicolumn{1}{c}{--} & 16.87 $\pm$ 0.02 &  8.18 $\pm$ 0.25 & -0.43 $\pm$ 0.02 &  2.446 $\pm$ 0.031 & 0.0954 $\pm$ 
	0.0031 & \multicolumn{1}{c}{--} \\
  192 & 10 & 0.0989 $\pm$ 0.0008 & 20.41 $\pm$ 0.02 &  8.81 $\pm$ 0.14 & -0.60 $\pm$ 0.01 &  1.611 $\pm$ 0.012 & 0.0892 $\pm$ 
	0.0007 & -0.930 $\pm$ 0.023 \\
  197 & 5 & \multicolumn{1}{c}{--} & 20.37 $\pm$ 0.02 &  9.51 $\pm$ 0.24 & -0.80 $\pm$ 0.02 &  2.531 $\pm$ 0.129 & 0.1371 $\pm$ 
	0.0059 & \multicolumn{1}{c}{--} \\
  214 & 7 & \multicolumn{1}{c}{--} & 14.91 $\pm$ 0.02 & 3.55 $\pm$ 0.34 & -0.44 $\pm$ 0.03 & 0.893 $\pm$ 0.020 & 0.0447 $\pm$ 
	0.0010 & \multicolumn{1}{c}{--} \\
  219 & 7 & 0.0926 $\pm$ 0.0065 & 19.71 $\pm$ 0.02 &  7.24 $\pm$ 0.14 & -0.76 $\pm$ 0.07 &  1.698 $\pm$ 0.050 & 0.0909 $\pm$ 
	0.0024 & -0.889 $\pm$ 0.043 \\
  234 & 15 & 0.1251 $\pm$ 0.0031 & 29.31 $\pm$ 0.02 & 12.59 $\pm$ 0.36 & -1.56 $\pm$ 0.04 & 1.617 $\pm$ 0.048 & 0.1603 $\pm$ 
	0.0045 & -0.596 $\pm$ 0.023 \\
  236 & 7 & \multicolumn{1}{c}{--} & 26.75 $\pm$ 0.02 & 12.66 $\pm$ 0.05 & -1.26 $\pm$ 0.01 & 1.805 $\pm$ 0.018 & 0.1688 $\pm$ 
	0.0010 & \multicolumn{1}{c}{--} \\
  335 & 10 & \multicolumn{1}{c}{--} & 16.47 $\pm$ 0.02 &  7.45 $\pm$ 0.15 & -1.14 $\pm$ 0.02 &  5.368 $\pm$ 0.061 & 0.2286 $\pm$ 
	0.0022 & \multicolumn{1}{c}{--} \\
  347 & 5 & \multicolumn{1}{c}{--} & 22.59 $\pm$ 0.02 & 10.15 $\pm$ 0.35 & -0.78 $\pm$ 0.01 &  1.769 $\pm$ 0.129 & 0.1128 $\pm$ 
	0.0091 & \multicolumn{1}{c}{--} \\
  376 & 5 & \multicolumn{1}{c}{--} & 20.43 $\pm$ 0.02 & 9.66 $\pm$ 0.37 & -0.47 $\pm$ 0.01 & 1.508 $\pm$ 0.017 & 0.0816 $\pm$ 
	0.0009 & \multicolumn{1}{c}{--} \\
  377 & 6 & \multicolumn{1}{c}{--} & 20.65 $\pm$ 0.02 & 9.69 $\pm$ 0.09 & -1.60 $\pm$ 0.02 & 4.957 $\pm$ 0.061 & 0.2738 $\pm$ 
	0.0028 & \multicolumn{1}{c}{--} \\
  387 & 9 & \multicolumn{1}{c}{--} & 28.51 $\pm$ 0.02 & 13.47 $\pm$ 0.13 & -1.41 $\pm$ 0.01 & 1.601 $\pm$ 0.061 & 0.1775 $\pm$ 
	0.0095 & \multicolumn{1}{c}{--} \\
  409 & 9 & \multicolumn{1}{c}{--} & 19.67 $\pm$ 0.02 & 8.96 $\pm$ 0.18 & -1.54 $\pm$ 0.02  & 5.049 $\pm$ 0.095 & 0.2629 $\pm$ 
	0.0030 & \multicolumn{1}{c}{--} \\
  419 & 6 & \multicolumn{1}{c}{--} & 13.99 $\pm$ 0.02 &  6.60 $\pm$ 0.11 & -1.27 $\pm$ 0.04 &  9.810 $\pm$ 0.291 & 0.3236 $\pm$ 
	0.0040 & \multicolumn{1}{c}{--} \\
  434 & 6 & \multicolumn{1}{c}{--} & 19.07 $\pm$ 0.02 &  7.70 $\pm$ 0.21 & -0.33 $\pm$ 0.04 &  0.909 $\pm$ 0.025 & 0.0476 $\pm$ 
	0.0015 & \multicolumn{1}{c}{--} \\
  472 & 7 & 0.1753 $\pm$ 0.0161 & 19.27 $\pm$ 0.02 & 9.29 $\pm$ 0.11 & -0.64 $\pm$ 0.01 & 2.564 $\pm$ 0.050 & 0.1243 $\pm$ 
	0.0017 & -0.714 $\pm$ 0.036 \\
  511 & 10 & 0.2829 $\pm$ 0.0100 & 19.59 $\pm$ 0.02 &  9.15 $\pm$ 0.07 & -1.66 $\pm$ 0.02 &  5.815 $\pm$ 0.083 & 0.2969 $\pm$ 
	0.0033 & -0.299 $\pm$ 0.018 \\
  584 & 6 & 0.1080 $\pm$ 0.0028 & 19.33 $\pm$ 0.02 &  7.53 $\pm$ 0.11 & -0.75 $\pm$ 0.04 &  1.892 $\pm$ 0.027 & 0.1004 $\pm$ 
	0.0015 & -0.838 $\pm$ 0.027 \\
  660 & 9 & \multicolumn{1}{c}{--} & 18.81 $\pm$ 0.02 &  6.51 $\pm$ 0.60 & -0.63 $\pm$ 0.05 &  1.400 $\pm$ 0.074 & 0.0729 $\pm$ 
	0.0038 & \multicolumn{1}{c}{--} \\
  679 & 7 & 0.1825 $\pm$ 0.0046 & 27.23 $\pm$ 0.90 & \multicolumn{1}{c}{--} & \multicolumn{1}{c}{--} & \multicolumn{1}{c}{--} &
	\multicolumn{1}{c}{--} & \multicolumn{1}{c}{--} \\
  796 & 7 & 0.1145 $\pm$ 0.0032 & 21.07 $\pm$ 0.02 &  6.71 $\pm$ 0.33 & -1.21 $\pm$ 0.10 &  2.122 $\pm$ 0.062 & 0.1132 $\pm$ 
	0.0017 & -0.692 $\pm$ 0.034 \\
  863 & 7 & \multicolumn{1}{c}{--} & 18.25 $\pm$ 0.02 & 8.95 $\pm$ 0.13 & -0.19 $\pm$ 0.01 & 0.936 $\pm$ 0.038 & 0.0405 $\pm$ 
	0.0014 & \multicolumn{1}{c}{--} \\
  980 & 6 & \multicolumn{1}{c}{--} & 29.11 $\pm$ 0.02 & 12.20 $\pm$ 0.14 & -1.24 $\pm$ 0.01 &  1.317 $\pm$ 0.049 & 0.1231 $\pm$ 
	0.0068 & \multicolumn{1}{c}{--} \\
 1021 & 10 & \multicolumn{1}{c}{--} & 15.63 $\pm$ 0.02 &  7.51 $\pm$ 0.10 & -0.71 $\pm$ 0.02 &  4.673 $\pm$ 0.042 & 0.1675 $\pm$ 
        0.0017 & \multicolumn{1}{c}{--} \\
 2867 & 6 & \multicolumn{1}{c}{--} & 17.15 $\pm$ 0.02 & 4.11 $\pm$ 0.62 & -0.41 $\pm$ 0.04 & 0.737 $\pm$ 0.020 & 0.0370 $\pm$ 
        0.0010 & \multicolumn{1}{c}{--} \\
      &                   &                &                &                  &                  &                   & 
\end{tabular}
\end{table*}			
%
\begin{table*} 
\label{Tab:Tab2}
\centering
\caption{Geometric albedo values $p_V$ for all asteroids not belonging to the Shevchenko and Tedesco (2006) 
list, for which we have polarimetric observations suited to derive the albedo using one or more of the 
relations explained in Paper I. The columns marked as ``no low-$p_V$'' refer to calibrations of the 
$h$ - $p_V$ and $P_{\rm min}$ - $p_V$ relations computed using asteroids having $p_V > 0.08$, only (see text).}  
\tiny
\begin{tabular}{rcccccc}
 & & & \\
\multicolumn{1}{c}{Number} & $p_V(h)$ & $p_V(h)$ & $p_V(h_{ABC})$ & $p_V(P_{\rm min})$ & $p_V(\Psi)$ & $p_V(p\ast)$ \\
                           &    & (no low-$p_V$) &             & (no low-$p_V$)    &             &              \\
 & & & \\
    5 & 0.226 $\pm$ 0.023 & 0.224 $\pm$ 0.030 & 0.206 $\pm$ 0.016 & 0.219 $\pm$ 0.007 & 0.200 $\pm$ 0.007 & 0.222 $\pm$ 0.020 \\
    6 & 0.228 $\pm$ 0.025 & 0.225 $\pm$ 0.030 & 0.202 $\pm$ 0.016 & 0.195 $\pm$ 0.005 & 0.214 $\pm$ 0.007 & 0.208 $\pm$ 0.019 \\
    7 & 0.193 $\pm$ 0.018 & 0.200 $\pm$ 0.025 & 0.189 $\pm$ 0.015 & 0.208 $\pm$ 0.025 & 0.199 $\pm$ 0.007 & 0.195 $\pm$ 0.016 \\
    9 & 0.328 $\pm$ 0.047 & 0.293 $\pm$ 0.046 & 0.286 $\pm$ 0.027 & 0.220 $\pm$ 0.006 & 0.289 $\pm$ 0.013 & 0.283 $\pm$ 0.032 \\
   10 & \multicolumn{1}{c}{--} & \multicolumn{1}{c}{--} & 0.075 $\pm$ 0.005 & \multicolumn{1}{c}{--} & 0.078 $\pm$ 0.004 
& \multicolumn{1}{c}{--} \\
   11 & 0.176 $\pm$ 0.017 & 0.187 $\pm$ 0.023 & 0.171 $\pm$ 0.013 & 0.208 $\pm$ 0.008 & 0.163 $\pm$ 0.006 & 0.184 $\pm$ 0.015 \\
   12 & 0.159 $\pm$ 0.014 & 0.174 $\pm$ 0.021 & 0.145 $\pm$ 0.010 & 0.197 $\pm$ 0.004 & 0.183 $\pm$ 0.006 & 0.167 $\pm$ 0.013 \\
   13 & 0.072 $\pm$ 0.005 & \multicolumn{1}{c}{--} & 0.059 $\pm$ 0.004 & \multicolumn{1}{c}{--} & 0.073 $\pm$ 0.004 & 
  0.060 $\pm$ 0.003 \\
   14 & 0.584 $\pm$ 0.087 & 0.444 $\pm$ 0.076 & \multicolumn{1}{c}{--} & \multicolumn{1}{c}{--} & \multicolumn{1}{c}{--} & 
  \multicolumn{1}{c}{--} \\
   15 & \multicolumn{1}{c}{--} & \multicolumn{1}{c}{--} & 0.187 $\pm$ 0.021 & 0.219 $\pm$ 0.008 & 0.191 $\pm$ 0.016 & 
  \multicolumn{1}{c}{--} \\
   16 & 0.173 $\pm$ 0.016 & 0.185 $\pm$ 0.023 & 0.173 $\pm$ 0.013 & \multicolumn{1}{c}{--} & 0.186 $\pm$ 0.006 & 
  0.154 $\pm$ 0.012 \\
   18 & 0.202 $\pm$ 0.019 & 0.207 $\pm$ 0.026 & 0.187 $\pm$ 0.014 & 0.194 $\pm$ 0.004 & 0.199 $\pm$ 0.007 & 0.192 $\pm$ 0.016 \\
   19 & 0.079 $\pm$ 0.006 & 0.105 $\pm$ 0.011 & 0.059 $\pm$ 0.003 & \multicolumn{1}{c}{--} & 0.073 $\pm$ 0.003 & 
  0.069 $\pm$ 0.004 \\
   20 & 0.285 $\pm$ 0.047 & 0.265 $\pm$ 0.044 & 0.301 $\pm$ 0.026 & 0.225 $\pm$ 0.008 & 0.266 $\pm$ 0.009 & 0.263 $\pm$ 0.033 \\
   21 & 0.163 $\pm$ 0.014 & 0.177 $\pm$ 0.021 & 0.146 $\pm$ 0.011 & \multicolumn{1}{c}{--} & 0.181 $\pm$ 0.006 & 
  0.131 $\pm$ 0.009 \\
   24 & \multicolumn{1}{c}{--} & \multicolumn{1}{c}{--} & 0.069 $\pm$ 0.004 & \multicolumn{1}{c}{--} & 0.078 $\pm$ 0.004 & 
  \multicolumn{1}{c}{--} \\
   29 & \multicolumn{1}{c}{--} & \multicolumn{1}{c}{--} & 0.168 $\pm$ 0.013 & 0.183 $\pm$ 0.004 & 0.185 $\pm$ 0.007 & 
  \multicolumn{1}{c}{--} \\
   30 & 0.227 $\pm$ 0.022 & 0.225 $\pm$ 0.029 & 0.196 $\pm$ 0.015 & 0.201 $\pm$ 0.006 & 0.197 $\pm$ 0.007 & 0.211 $\pm$ 0.018 \\
   40 & 0.248 $\pm$ 0.033 & 0.240 $\pm$ 0.036 & 0.234 $\pm$ 0.021 & 0.191 $\pm$ 0.007 & 0.224 $\pm$ 0.010 & 0.217 $\pm$ 0.023 \\
   42 & 0.181 $\pm$ 0.019 & 0.191 $\pm$ 0.025 & 0.191 $\pm$ 0.015 & 0.221 $\pm$ 0.009 & 0.174 $\pm$ 0.006 & 0.194 $\pm$ 0.017 \\
   43 & 0.296 $\pm$ 0.031 & 0.272 $\pm$ 0.038 & \multicolumn{1}{c}{--} & \multicolumn{1}{c}{--} & \multicolumn{1}{c}{--} & 
  \multicolumn{1}{c}{--} \\
   44 & 0.886 $\pm$ 0.177 & 0.599 $\pm$ 0.122 & 0.965 $\pm$ 0.108 & 0.412 $\pm$ 0.024 & 0.704 $\pm$ 0.030 & 0.791 $\pm$ 0.013 \\
   54 & \multicolumn{1}{c}{--} & \multicolumn{1}{c}{--} & 0.058 $\pm$ 0.003 & \multicolumn{1}{c}{--} & 0.073 $\pm$ 0.003 & 
  \multicolumn{1}{c}{--} \\
   56 & 0.070 $\pm$ 0.005 & \multicolumn{1}{c}{--} & 0.064 $\pm$ 0.004 & \multicolumn{1}{c}{--} & 0.067 $\pm$ 0.003 & 
  0.071 $\pm$ 0.004 \\
   63 & \multicolumn{1}{c}{--} & \multicolumn{1}{c}{--} & 0.214 $\pm$ 0.020 & 0.188 $\pm$ 0.019 & 0.197 $\pm$ 0.010 & 
  \multicolumn{1}{c}{--} \\
   71 & \multicolumn{1}{c}{--} & \multicolumn{1}{c}{--} & 0.357 $\pm$ 0.032 & 0.234 $\pm$ 0.022 & 0.298 $\pm$ 0.010 & 
  \multicolumn{1}{c}{--} \\
   77 & \multicolumn{1}{c}{--} & \multicolumn{1}{c}{--} & 0.210 $\pm$ 0.041 & \multicolumn{1}{c}{--} & 0.205 $\pm$ 0.021 & 
  \multicolumn{1}{c}{--} \\
   84 & \multicolumn{1}{c}{--} & \multicolumn{1}{c}{--} & 0.066 $\pm$ 0.004 & \multicolumn{1}{c}{--} & 0.078 $\pm$ 0.004 & 
  \multicolumn{1}{c}{--} \\
   87 & \multicolumn{1}{c}{--} & \multicolumn{1}{c}{--} & 0.103 $\pm$ 0.007 & \multicolumn{1}{c}{--} & 0.117 $\pm$ 0.005 & 
  \multicolumn{1}{c}{--} \\
   92 & \multicolumn{1}{c}{--} & \multicolumn{1}{c}{--} & 0.329 $\pm$ 0.029 & 0.201 $\pm$ 0.006 & 0.304 $\pm$ 0.010 & 
  \multicolumn{1}{c}{--} \\
   97 & 0.120 $\pm$ 0.011 & 0.142 $\pm$ 0.016 & 0.087 $\pm$ 0.006 & \multicolumn{1}{c}{--} & 0.109 $\pm$ 0.004 & 
  0.109 $\pm$ 0.008 \\
  113 & 0.289 $\pm$ 0.041 & 0.268 $\pm$ 0.041 & 0.276 $\pm$ 0.024 & 0.286 $\pm$ 0.018 & 0.245 $\pm$ 0.009 & 0.308 $\pm$ 0.037 \\
  115 & \multicolumn{1}{c}{--} & \multicolumn{1}{c}{--} & 0.208 $\pm$ 0.017 & 0.222 $\pm$ 0.006 & 0.213 $\pm$ 0.007 & 
 \multicolumn{1}{c}{--} \\
  128 & \multicolumn{1}{c}{--} & \multicolumn{1}{c}{--} & 0.064 $\pm$ 0.004 & \multicolumn{1}{c}{--} & 0.070 $\pm$ 0.004 & 
 \multicolumn{1}{c}{--} \\
  131 & \multicolumn{1}{c}{--} & \multicolumn{1}{c}{--} & 0.244 $\pm$ 0.021 & 0.322 $\pm$ 0.015 & 0.236 $\pm$ 0.009 & 
 \multicolumn{1}{c}{--} \\
  132 & 0.156 $\pm$ 0.014 & 0.172 $\pm$ 0.021 & 0.146 $\pm$ 0.011 & \multicolumn{1}{c}{--} & 0.143 $\pm$ 0.006 & 
 0.136 $\pm$ 0.011 \\
  138 & 0.200 $\pm$ 0.030 & 0.205 $\pm$ 0.031 & \multicolumn{1}{c}{--} & \multicolumn{1}{c}{--} & \multicolumn{1}{c}{--} & 
 \multicolumn{1}{c}{--} \\
  161 & \multicolumn{1}{c}{--} & \multicolumn{1}{c}{--} & 0.109 $\pm$ 0.007 & \multicolumn{1}{c}{--} & 0.113 $\pm$ 0.005 & 
 \multicolumn{1}{c}{--} \\
  172 & 0.179 $\pm$ 0.017 & 0.189 $\pm$ 0.024 & 0.106 $\pm$ 0.009 & \multicolumn{1}{c}{--} & 0.206 $\pm$ 0.007 & 
 0.131 $\pm$ 0.010 \\
  186 & \multicolumn{1}{c}{--} & \multicolumn{1}{c}{--} & 0.205 $\pm$ 0.018 & 0.341 $\pm$ 0.022 & 0.144 $\pm$ 0.005 & 
\multicolumn{1}{c}{--} \\
  192 & 0.216 $\pm$ 0.020 & 0.217 $\pm$ 0.028 & 0.222 $\pm$ 0.018 & 0.254 $\pm$ 0.009 & 0.218 $\pm$ 0.007 & 0.238 $\pm$ 0.021 \\
  197 & \multicolumn{1}{c}{--} & \multicolumn{1}{c}{--} & 0.136 $\pm$ 0.012 & 0.198 $\pm$ 0.006 & 0.139 $\pm$ 0.009 & 
 \multicolumn{1}{c}{--} \\
  214 & \multicolumn{1}{c}{--} & \multicolumn{1}{c}{--} & 0.487 $\pm$ 0.047 & 0.334 $\pm$ 0.022 & 0.390 $\pm$ 0.015 & 
 \multicolumn{1}{c}{--} \\
  219 & 0.233 $\pm$ 0.028 & 0.229 $\pm$ 0.032 & 0.217 $\pm$ 0.018 & 0.207 $\pm$ 0.016 & 0.207 $\pm$ 0.009 & 0.219 $\pm$ 0.024 \\
  234 & 0.167 $\pm$ 0.015 & 0.180 $\pm$ 0.022 & 0.114 $\pm$ 0.009 & \multicolumn{1}{c}{--} & 0.217 $\pm$ 0.009 & 
 0.119 $\pm$ 0.009 \\
  236 & \multicolumn{1}{c}{--} & \multicolumn{1}{c}{--} & 0.107 $\pm$ 0.007 & \multicolumn{1}{c}{--} & 0.194 $\pm$ 0.007 & 
 \multicolumn{1}{c}{--} \\
  335 & \multicolumn{1}{c}{--} & \multicolumn{1}{c}{--} & 0.076 $\pm$ 0.005 & \multicolumn{1}{c}{--} & 0.066 $\pm$ 0.003 & 
 \multicolumn{1}{c}{--} \\
  347 & \multicolumn{1}{c}{--} & \multicolumn{1}{c}{--} & 0.170 $\pm$ 0.020 & 0.201 $\pm$ 0.005 & 0.198 $\pm$ 0.016 & 
 \multicolumn{1}{c}{--} \\
  376 & \multicolumn{1}{c}{--} & \multicolumn{1}{c}{--} & 0.245 $\pm$ 0.020 & 0.316 $\pm$ 0.013 & 0.232 $\pm$ 0.008 & 
 \multicolumn{1}{c}{--} \\
  377 & \multicolumn{1}{c}{--} & \multicolumn{1}{c}{--} & 0.062 $\pm$ 0.004 & \multicolumn{1}{c}{--} & 0.072 $\pm$ 0.003 & 
 \multicolumn{1}{c}{--} \\
  387 & \multicolumn{1}{c}{--} & \multicolumn{1}{c}{--} & 0.101 $\pm$ 0.009 & \multicolumn{1}{c}{--} & 0.219 $\pm$ 0.011 & 
 \multicolumn{1}{c}{--} \\
  409 & \multicolumn{1}{c}{--} & \multicolumn{1}{c}{--} & 0.065 $\pm$ 0.004 & \multicolumn{1}{c}{--} & 0.070 $\pm$ 0.004 & 
 \multicolumn{1}{c}{--} \\
  419 & \multicolumn{1}{c}{--} & \multicolumn{1}{c}{--} & 0.051 $\pm$ 0.003 & \multicolumn{1}{c}{--} & 0.037 $\pm$ 0.002 & 
 \multicolumn{1}{c}{--} \\
  434 & \multicolumn{1}{c}{--} & \multicolumn{1}{c}{--} & 0.453 $\pm$ 0.045 & 0.423 $\pm$ 0.044 & 0.383 $\pm$ 0.015 & 
 \multicolumn{1}{c}{--} \\
  472 & 0.115 $\pm$ 0.015 & 0.137 $\pm$ 0.018 & 0.152 $\pm$ 0.011 & 0.238 $\pm$ 0.007 & 0.138 $\pm$ 0.006 & 0.152 $\pm$ 0.015 \\
  511 & 0.067 $\pm$ 0.005 & \multicolumn{1}{c}{--} & 0.056 $\pm$ 0.003 & \multicolumn{1}{c}{--} & 0.061 $\pm$ 0.003 & 
 0.065 $\pm$ 0.004 \\
  584 & 0.196 $\pm$ 0.019 & 0.202 $\pm$ 0.026 & 0.194 $\pm$ 0.015 & 0.208 $\pm$ 0.010 & 0.186 $\pm$ 0.007 & 0.197 $\pm$ 0.017 \\
  660 & \multicolumn{1}{c}{--} & \multicolumn{1}{c}{--} & 0.279 $\pm$ 0.029 & 0.242 $\pm$ 0.017 & 0.250 $\pm$ 0.015 & 
 \multicolumn{1}{c}{--} \\
  679 & 0.110 $\pm$ 0.009 & 0.133 $\pm$ 0.015 & \multicolumn{1}{c}{--} & \multicolumn{1}{c}{--} & \multicolumn{1}{c}{--} & 
 \multicolumn{1}{c}{--} \\
  796 & 0.184 $\pm$ 0.017 & 0.193 $\pm$ 0.024 & 0.169 $\pm$ 0.013 & \multicolumn{1}{c}{--} & 0.166 $\pm$ 0.007 & 
 0.146 $\pm$ 0.013 \\
  863 & \multicolumn{1}{c}{--} & \multicolumn{1}{c}{--} & 0.545 $\pm$ 0.057 & 0.682 $\pm$ 0.053 & 0.372 $\pm$  0.019 & 
 \multicolumn{1}{c}{--} \\
  980 & \multicolumn{1}{c}{--} & \multicolumn{1}{c}{--} & 0.154 $\pm$ 0.015 & \multicolumn{1}{c}{--} & 0.265 $\pm$ 0.013 & 
 \multicolumn{1}{c}{--} \\
 1021 & \multicolumn{1}{c}{--} & \multicolumn{1}{c}{--} & 0.108 $\pm$ 0.007 & 0.219 $\pm$ 0.008 & 0.076 $\pm$ 0.004 & 
 \multicolumn{1}{c}{--} \\
 2867 & \multicolumn{1}{c}{--} & \multicolumn{1}{c}{--} & 0.604 $\pm$ 0.062 & 0.350 $\pm$ 0.033 & 0.471 $\pm$ 0.019 & 
 \multicolumn{1}{c}{--} \\
 & & &
\end{tabular}
\end{table*}
The difference between the polarization slopes $h$ and $h_{ABC}$
consists only in the way they are computed.
In Paper I, we showed that the use of $h$ or $h_{ABC}$ gives very
similar solutions, apart from marginally better RMS deviations in the case of 
using $h$. So, in practical situations the use of either $h$ or $h_{ABC}$ is mainly 
dictated by the available polarimetric data. 

It is important to note that in Paper I we considered two kinds of calibration 
of the slope - albedo and $P_{\rm min}$ - albedo relations.
The first was obtained by fitting all data available for all the
objects of our dataset. The second was obtained by removing
from the analysis asteroids that according to \citet{ShevTed} have
albedos lower than $0.08$. This was suggested by the evidence that
the slope - albedo and (even more) the $P_{\rm min}$ - albedo relation 
tend to saturate at low-albedo. We showed that the
linear best-fit solutions obtained by exluding low-albedo asteroids from
the analysis give smaller RMS deviations of the data, and should
therefore be preferred, but only in narrower intervals of $h$ and 
$P_{\rm min}$, because at values of $h > 0.25$\%/$^\circ$ and $P_{\rm min}$ deeper
than $-1$\%, there is an unsolvable ambiguity between objects of
quite different albedo, but sharing the same values of $h$ (or $h_{ABC}$) 
and $P_{\rm min}$. We found also that the slope - albedo relation calibrated
against asteroids of all albedos can still be used to derive a decent
albedo estimate for any object, whereas the use of the  
$P_{\rm min}$ - albedo relation calibrated against asteroids of all albedos 
should not be used, since the resulting errors on the derived albedos
are exceedingly high.

In Table 2 we list the albedo values obtained from the values of different
polarimetric parameters considered in Paper I. We give, whenever
possible, two values of albedo obtained from the
$h$ slope, one corresponding to the calibration of the slope - albedo 
relation using all calibration asteroids considered in Paper I, and one
corresponding to the calibration obtained considering only the objects having 
albedo larger than $0.08$. The latter albedo value is given only 
for asteroids having $h < 0.25\%/^\circ$. As for the
albedo computed using $P_{\rm min}$, we use only the calibration obtained
in Paper I for asteroids having albedo larger than $0.08$, and we list
the corresponding albedo only for asteroids having $P_{\rm min}$
not reaching $1$\%. In so doing, we are following our own recommendations 
as explained in Paper I. In the case of $h_{ABC}$ and $\Psi$, the resulting
albedo values given in Table 2 are 
based on the whole set of calibration asteroids, including also low-albedo objects.
In Paper I we showed that in the case of $\Psi$-based albedos, the resulting
values are generally very reliable. In the case of $h_{ABC}$, however,
it should be better to exclude from the calibration the asteroids having
albedo smaller than $0.08$ (as in the case of $h$ just mentioned). The 
albedo values listed in Table 2 corresponding to $h_{ABC}$ must therefore
be taken with some caveat, since some overestimation of albedos, specially
for high-albedo objects, is likely present.

\begin{figure}
\begin{center}
\includegraphics[width=88mm]{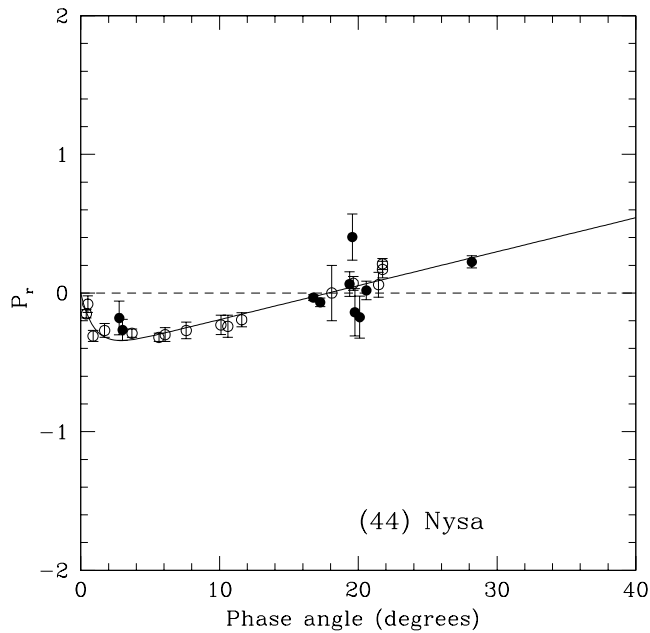}
\end{center}
 \caption{The phase - polarization curve of the high-albedo asteroid (44) Nysa.
In this and other figures, full symbols indicate measurements taken  
at the CASLEO observatory and recently published by \citet{Giletal14}. 
Open symbols are other measurements taken from the literature. The solid curve is the
best-fit curve using the exponential - linear representation (Eq.$\:$\ref{Eqn:ABC}).}
 \label{Fig:Nysa}
\end{figure}
From Table 2, we can see that most albedo values are obtained from 
computation of the $h_{ABC}$ and $\Psi$ polarimetric parameters, 
obtained from a fit of the phase - polarization curve using
Eq.$~$\ref{Eqn:ABC}. 
In general, the agreement among the albedos
obtained using different polarimetric parameters
is quite good. This confirms that, depending on the available data,
reliable albedo values can be obtained even from
a fairly small number of polarimetric measurements. The
best polarimetric parameter to be used depends case by case upon
the available data, and we have outlined in Paper I 
how to proceed in practical situations. 

The albedo values obtained from the $\Psi$ parameter
tend to vary in a more limited interval with respect to
what we find using other polarimetric parameters. This is
particularly true for the highest-albedo objects,
(44) Nysa, (214) Aschera, and (2867) Steins, included in our
sample. These asteroids belong to the old $E$-class defined by Tholen 
\citep{TholenBar}; in the more recent classification by \citet{BusBinzel} 
they are classified as $Xc$. There is some problem concerning  
in particular (44) Nysa. As shown in Table 2, we
find extremely high values, up to $0.9$ or even above, using the slope -
albedo relation calibrated against all asteroids considered in Paper I,
regardless of their albedo. By using calibrations obtained by dropping
low-albedo asteroids, the resulting albedo tends to decrease down to $0.6$. 
This  was one of the main reasons in Paper I to compute alternative 
calibrations based on the exclusion of the darkest calibration objects.
In the case of the $p*$ - albedo relation, we obtain for Nysa a still
high value around $0.8$. A slightly more moderate value, around $0.7$, is 
found using the $\Psi$ parameter. The extremely shallow polarimetric slope 
of Nysa is shown in Fig.$\:$\ref{Fig:Nysa}. In this figure, note also the 
very low value of the phase angle corresponding to $P_{\rm min}$.
The albedos of (214) Aschera and (2867) Steins, 
turn out to be much lower than in the case of (44) Nysa. 
Also for another $E$-class asteroid, 
(434) Hungaria, that is now classified as $Xe$ by \citet{BusBinzel}, we 
find a much more moderate albedo value, slightly above $0.4$. In general terms,
$E$-class asteroids are those for which the choice of the polarimetric parameter 
chosen to derive the albedo makes the most difference, and for which it is highly
recommended to use calibrations of the polarimetric slope and of $P_{\rm min}$ 
that are computed by excluding low-albedo objects from the computation. 
In other words, albedo values in columns 2 and 4 of Table 2
should not be used for $E$-class objects. In this way, with the notable
exception of (44) Nysa, the albedo values obtained are generally in
reasonable mutual agreement, ranging approximately between $0.4$ and $0.5$,
a range that one might expect corresponds to a real variation among the objects
of this class.

It is also interesting to note in Table 2 the high albedo values
found for the $A$-class asteroid (863) Benkoela, ranging from $0.4$ to 
$0.7$, depending on the adopted polarimetric parameter. This is the only
example of $A$-class objects in our sample. Further observations of
other members of this fairly rare class, which is thought to have
a composition dominated by olivine, are needed to confirm this 
preliminary result.  

When looking at the data displayed in Table 2, one should take into
account that this Table lists also a number of so-called Barbarian
objects, which are known to exhibit peculiar polarimetric properties,
and in particular a very wide width of the negative polarization
branch, up to about $30^\circ$ in phase
\citep{Cellinoetal06,Giletal08,MasCel09}.  For these objects, it is
likely that the peculiar morphology of the phase - polarization curve
can prevent us from deriving the albedo using the same polarimetric
parameters developed for normal asteroids. Asteroids (234) Barbara
(the prototype of the Barbarian class), (172) Baucis, (236) Honoria,
(387) Aquitania and (980) Anacostia, included in Table 2, are all
Barbarians. For them, we see significant differences in the albedo values
derived using different polarimetric parameters. The albedos derived 
using the $\Psi$ parameter, tend to have values
around $0.20$, whereas in the case of using $h_{ABC}$ the corresponding
value tends to be around $0.10$. In both cases, (980) Anacostia
seems to have a higher albedo than the other Barbarians in our sample.

If we exclude Barbarians, we see that the albedo values
derived using different parameters show a remarkable consistency. We
conclude therefore that it may be risky to try and derive albedos for
Barbarian objects using the conventional polarimetric parameters, due
to their unusual polarimetric behaviour. The situation is much more
promising for the vast majority of ``normal'' asteroids.

\section{The distribution of polarimetric parameters}\label{distributions}
The availability of a data set of phase - polarization curves of
sufficiently good quality to obtain best-fit representations using the
exponential-linear relation also allows us to make some updated
investigations of the polarimetric behavior of different classes of
asteroids, not limited to the determination of the geometric albedo.

\begin{figure}
\begin{center}
\includegraphics[width=88mm]{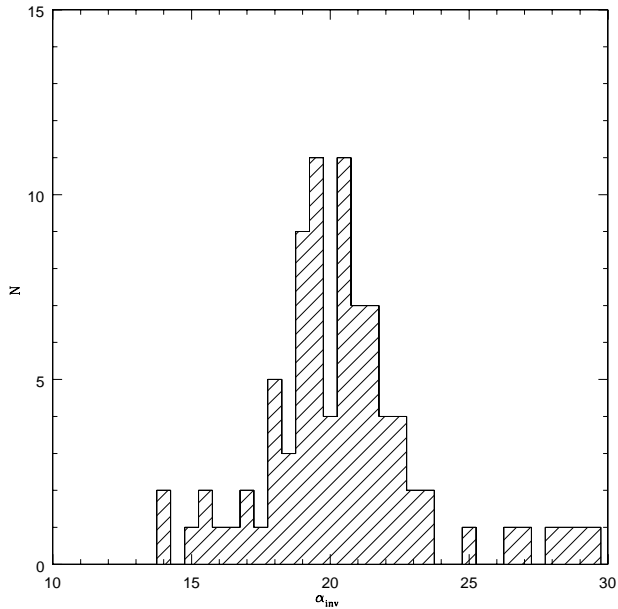}
\end{center}
 \caption{Histogram of the inversion angle $\alpha_{\rm inv}$ for
all asteroids considered in this paper and in Paper I.}
 \label{Fig:istalfinv}
\end{figure}
%
\begin{figure}
\begin{center}
\includegraphics[width=88mm]{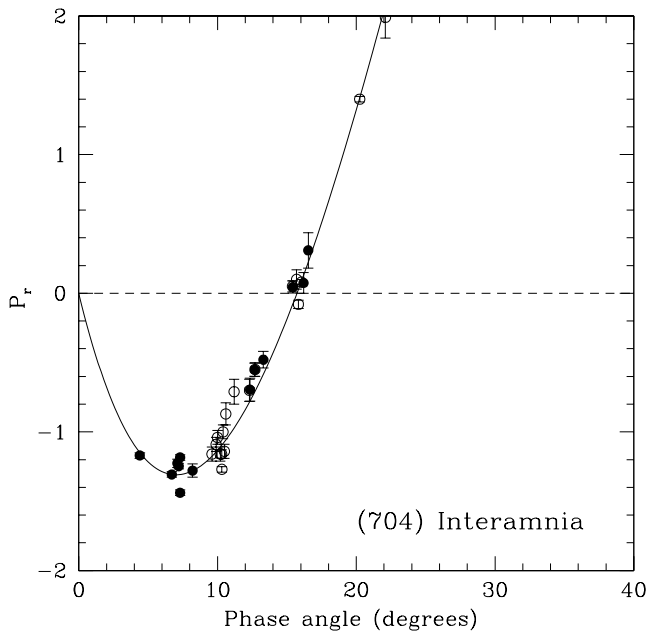}
\end{center}
 \caption{The phase - polarization curve of the low-albedo asteroid (704) Interamnia. 
Full symbols indicate measurements taken at the CASLEO observatory. Open symbols 
are other measurements taken from the literature. The solid curve is the
best-fit curve using the exponential - linear representation (Eq.$\:$\ref{Eqn:ABC}).}
 \label{Fig:Interamnia}
\end{figure}
\begin{figure}
\begin{center}
\includegraphics[width=88mm]{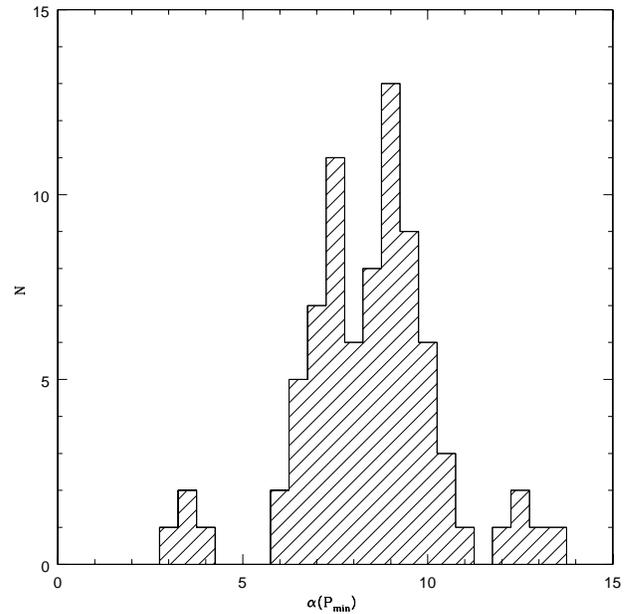}
\end{center}
 \caption{Histogram of the values of phase angle corresponding 
to $P_{\rm min}$  for all asteroids conidered in this paper and in Paper I.}
 \label{Fig:istphasepmin}
\end{figure}
%
In particular, we can analyze the distributions of the 
inversion angle $\alpha_{\rm inv}$, and of the phase angle corresponding to
$P_{\rm min}$, as shown in Fig.$~$\ref{Fig:istalfinv} and \ref{Fig:istphasepmin},
respectively.

Since the early days of asteroid polarimetry it has been known that the inversion angle 
occurs at phase angles close to $20^\circ$. In recent years, however, some important 
exceptions have been found, as shown in  Fig.$~$\ref{Fig:istalfinv}. By looking at 
the low-end distribution of $\alpha_{\rm inv}$, there are two objects characterized 
by an inversion angle around $14^\circ$. In the case of 
asteroid (14) Irene, we cannot draw any conclusion because its phase - polarization
curve is not of a very good quality and new observations are needed
to better understand its true behaviour. The case of (419) Aurelia, conversely, 
is much more interesting.
This asteroid belongs to the old $F$ taxonomic class identified 
by \citet{GradieTed82} \citep[see also][]{Tedescoetal89}. (419) Aurelia is no 
longer identified as an $F$ class
in more recent taxonomic classifications based on reflectance spectra that no longer cover
the blue part of the spectrum. Asteroids previously classified as $F$ are now included in
the modern $B$ class \citep{BusBinzel}. We know, however, that the original $F$-class 
asteroids can be distinguished based on their polarimetric properties. In particular,
these asteroids are characterized by small values of $\alpha_{\rm inv}$ \citep{Irina05}.
In addition to Aurelia, three other $F$-class asteroids in our sample show low 
values of $\alpha_{\rm inv}$, around $16^\circ$. They are (335) Roberta, (704) Interamnia 
(analyzed in Paper I), and (1021) Flammario, although in the case of Flammario the 
available polarimetric data are noisy. The excellent phase-polarization
curve of (704) Interamnia, one of the calibration asteroids used in Paper I, is
shown here in Fig.$~$\ref{Fig:Interamnia}.

In addition to the above-mentioned $F$-class asteroids, another asteroid, (214)
Aschera, exhibits an inversion angle of about $15^\circ$. This is a very high-albedo
asteroid, and its properties will be discussed below. A few other objects 
exhibit relatively low values of $\alpha_{\rm inv}$ around $18^\circ$. We have seen, 
however, that for them we still need additional observations to better cbaracterize their
phase - polarization curves.

At the other end of the distribution of $\alpha_{\rm inv}$, we see some objects
characterized by values well above $20^\circ$. These are the Barbarian asteroids 
already mentioned above. Four of them, (172) Baucis, (234) Barbara, (387) Aquitania,
and (980) Anacostia have inversion angles above $28^\circ$. (236) Honoria and (679) Pax 
have slightly lower inversion angles around $27^\circ$. The single object
exhibiting an inversion angle of about $25^\circ$ is (21) Lutetia. 
This object is one of the two asteroids observed by
the Rosetta probe (the other being (2867) Steins, also included in our Tables). 
Lutetia has fairly unusual properties. It was
classified in the past as an $M$-class, possibly metal-rich, asteroid 
\citep{TholenBar}, but the observations performed before and during the Rosetta fly-by,
using different techniques, have shown that the composition of this
asteroid seems to be more compatible with that of some classes of
primitive meteorites \citep{Angioletta11}. The high value of the inversion angle 
of (21) Lutetia confirms that the surface of this asteroid is unusual.

The distribution of the inversion angle of polarization among asteroids 
appears to be today much wider and interesting than in the past. The
$\alpha_{\rm inv}$ parameter appears to be useful to distinguish classes of asteroids
with unusual surface properties. In the case of Barbarians, 
there are reasons to believe that these objects might be the remnants of the
first generation of planetesimals accreted in the first few $10^5$ years
of our solar system's history \citep[see][and references therein]{Watsonia}.
In the case of objects exhibiting small values of $\alpha_{\rm inv}$, it is
interesting to note that this property, typical of asteroids belonging to
the old $F$ class of \citet{GradieTed82}, has been found to be shared also by a few
cometary nuclei \citep{Bagnulocomete}. This interesting result tends
to strengthen other pieces of evidence of a possible link between $F$-class 
asteroids and comets, already suggested by the fact that comet 
Wilson Hurrington was first discovered as an asteroid (numbered 4015), 
classified as $CF$ and another $F$-class asteroid, (3200) Phaeton, is 
known to be the source of the Geminid meteors 
\citep[][and references therein]{Chamberlinetal96}.

\begin{figure}
\begin{center}
\includegraphics[width=88mm]{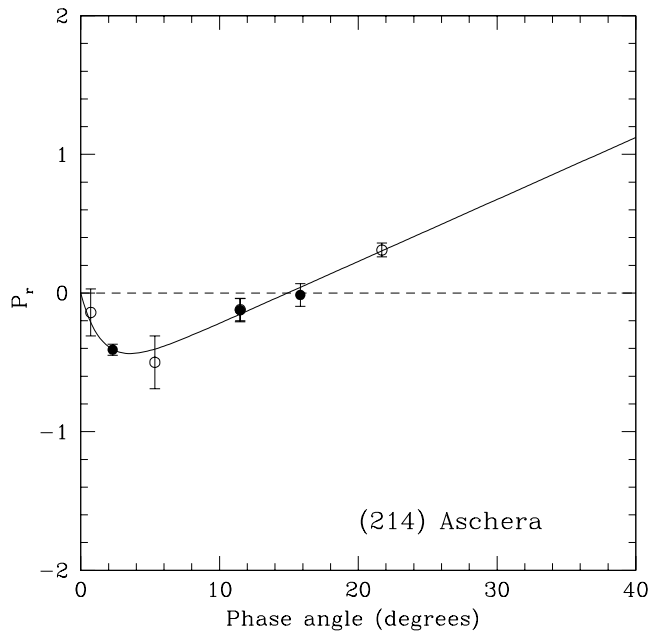}
\end{center}
 \caption{The phase - polarization curve of the high-albedo asteroid 
(214) Aschera. Full symbols indicate measurements taken at the CASLEO observatory. Open symbols 
are other measurements taken from the literature. Two different measurements obtained by different 
authors at a phase angle of $11.5^\circ$ perfectly overlap in this plot. The solid curve is the
best-fit curve using the exponential - linear representation (Eq.$\:$\ref{Eqn:ABC}).}
 \label{Fig:Aschera}
\end{figure}
The distribution of the phase angle corresponding to $P_{\rm min}$,
$\alpha(P_{\rm min})$, is shown in Fig.$\:$\ref{Fig:istphasepmin}.  We
see a confirmation of the fact that $P_{\rm min}$ is mostly found at
phase angles between $7^\circ$ and $10^\circ$. A few exceptions exist,
however. We find four asteroids having $\alpha(P_{\rm min})$ between $3^\circ$ and 
$4^\circ$, namely (44) Nysa, (71) Niobe, (214) Aschera and (2867) Steins.
We have already seen that (44), (214) and (2867) are high-albedo asteroids
belonging to the old $E$-class. As for (71) Niobe, it is classified as $Xe$
by \citet{BusBinzel}. Its albedo, according to our results listed in Table 2,
is high, though not as extreme as those of Nysa, Aschera, and Steins.
Its phase - polarization curve, however, does not include data at phase
angles less than about $7^\circ$, so the formal value of $\alpha(P_{\rm min})$ 
listed in Table 1 for this asteroid is still very uncertain, and 
could be considerably wrong. 
It seems, anyway, that a very low value of $\alpha(P_{\rm min})$ can be
a common feature among high-albedo asteroids, although some of them, like
the $Xe$-class (434) Phocaea and the $A$-class (863) Benkoela do not 
share this property. It is possible that for the highest-albedo
asteroids, having very shallow polarimetric slopes, determination 
of the inversion angle and the phase angle of $P_{\rm min}$ might turn
out to be more uncertain than the results presented here. We have already 
seen in Fig.$~$\ref{Fig:Nysa}, however, that the low value of $\alpha(P_{\rm min})$
for (44) Nysa, seems well defined, and does not appear to be affected by 
a possible polarization opposition effect. Though not so densely sampled, 
the same can be seen for the phase - polarization curve of (214) Aschera, 
shown in Fig.$~$\ref{Fig:Aschera}. We note that 
very low values of  $\alpha(P_{\rm min})$ are known also for two
Centaur objects, (2060) Chiron and (10199) Chariklo, and for the TNO object 
(5145) Pholus \citep{Bagnuloetal06, Belskayaetal10}. 
A fundamental difference with respect to high-albedo asteroids
is that for objects at high heliocentric distances $P_{\rm min}$ is much deeper,
and suggests low-albedo surfaces. 

At the other end of the $\alpha(P_{\rm min})$ distribution, there are 
five asteroids having phase angle of $P_{\rm min}$ between $12^\circ$ and $14^\circ$.
These are the five Barbarian asteroids discussed above. Having
a very wide negative polarization branch, it is not too surprising
that these objects tend also to have $P_{\rm min}$ at larger phase angles
than usual, and this feature might be related to the same surface
properties that determine the wide negative polarization branch. We note, 
however, that (21) Lutetia, which also
exhibits a large value of the inversion angle, is a perfectly 
normal asteroid as far as $\alpha(P_{\rm min})$ is concerned. 

\begin{figure}
\begin{center}
\includegraphics[width=88mm]{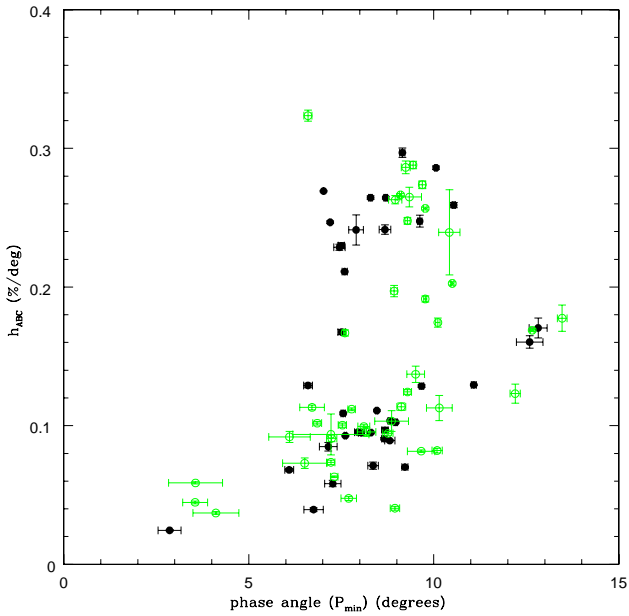}
\end{center}
 \caption{The $h_{ABC}$ polarimetric slope \emph{versus} phase angle of $P_{\rm min}$ 
for the whole sample of asteroids considered in Tables 1 and 2 and Paper I. 
Note that higher values of $h_{ABC}$ correspond to lower values of albedo. 
Objects having fewer than $10$ polarimetric measurements are indicated by open, 
green symbols.}
 \label{Fig:alfpminhABCall}
\end{figure}
\begin{figure}
\begin{center}
\includegraphics[width=88mm]{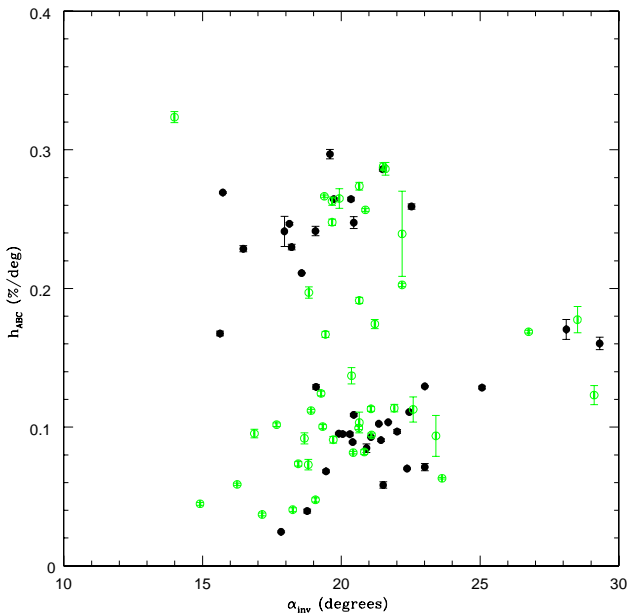}
\end{center}
 \caption{The $h_{ABC}$ polarimetric slope \emph{versus} inversion angle $\alpha_{inv}$  
 for the whole sample of asteroids considered in Table 1 and Paper I. Asteroids 
 for which we have fewer than $10$ polarimetric measurements are indicated by open, 
 green symbols.}
 \label{Fig:alfinvhABCall}
\end{figure}
%
\section{Relations between polarimetric parameters} \label{relations}

In Fig.$\:$\ref{Fig:alfpminhABCall} we present an 
$\alpha(P_{\rm min})$ - $h_{ABC}$ plot for all the asteroids
considered in this paper (Table 1) and including also those considered 
in Paper I. 
The meaning of this Figure is clear if we consider that the polarimetric 
slope ($h$, or, as in this case, $h_{ABC}$, to consider a larger number 
of objects) is diagnostic of the albedo. In particular, low-albedo
asteroids have higher values of $h_{ABC}$, and {\em viceversa}.

From the Figure, we see that the objects tend to split into two groups,
characterized by different average values of slope (i.e., albedo). 
This is expected considering that the main belt population is dominated
by two superclasses, namely the moderate-albedo $S$-class, and the
low-albedo $C$-class. 

In the lower part of the plot shown in Fig.$\:$\ref{Fig:alfpminhABCall},
corresponding to moderate to high-albedo objects, we see a general trend
of decreasing albedo for increasing $\alpha(P_{\rm min})$. A similar
trend, but slightly less pronounced, may be seen also in the upper part of the
plot. Low-albedo objects (having higher values of the polarimetric
slope $h_{ABC}$), tend to display values of $\alpha(P_{\rm min})$
which look more confined. 
Asteroids having $\alpha(P_{\rm min})$ below $6^\circ$ are found only 
among high-albedo objects, as already seen in Section$~$\ref{distributions}.
Most asteroids in Fig.$\:$\ref{Fig:alfpminhABCall} are located in the 
region between about $7^\circ$ and $10^\circ$ of $\alpha(P_{\rm  min})$. 
The right part of the plot is occupied by Barbarian asteroids, which tend 
to cluster at values of $h_{ABC}$ intermediate between those of low- and 
high-albedo objects.

In Fig.$\:$\ref{Fig:alfinvhABCall} we show a plot of 
the $h_{ABC}$ polarimetric slope \emph{versus} the inversion
angle $\alpha_{\rm inv}$. Again, we can see a clear splitting between low-albedo
asteroids and the rest of the population. In both groups, there is 
a trend for an increase of $h_{ABC}$ (equivalent to a decrease of the albedo) 
for increasing $\alpha_{\rm inv}$. This behaviour is possibly less sharp among
low-albedo objects, and $F$-class asteroids clearly do not
follow this trend, with (419) Aurelia occupying the most extreme top-left
location in the plot. 
All asteroids, independent of their albedo, tend to share the same interval
of inversion angles, except for the Barbarians, which occupy the high-end of
the  $\alpha_{\rm inv}$ range. Interesting enough, the location of the 
Barbarians in this plot tends to correspond to an extrapolation towards
larger values of  $\alpha_{\rm inv}$ of the relation between  $\alpha_{\rm inv}$
and $h_{ABC}$ exhibited by intermediate to high-albedo asteroids. We note again
also the location of (21) Lutetia, that with an $\alpha_{\rm inv}$ angle of
$25^\circ$, lies in between Barbarians and ``normal'' asteroids. 

\begin{figure}
\begin{center}
\includegraphics[width=88mm]{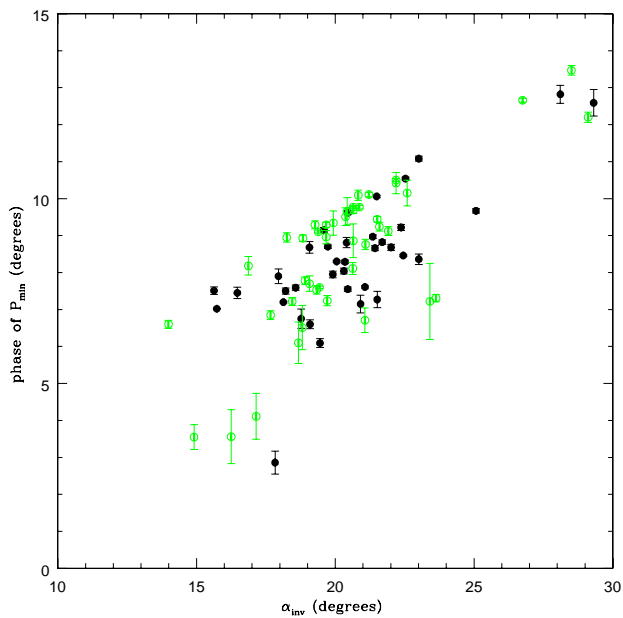}
\end{center}
 \caption{The $\alpha_{\rm inv}$ inversion angle \emph{versus} phase angle of $P_{\rm min}$ 
for the whole sample of asteroids considered in Table 1 and Paper I.  
Objects having fewer than $10$ polarimetric measurements are indicated by open, 
green symbols.}
 \label{Fig:alfpminalfinvall}
\end{figure}

Figs.$~$\ref{Fig:alfpminhABCall} and \ref{Fig:alfinvhABCall} suggest
a correlation between inversion angle and phase angle of $P_{\rm min}$ which
is clearly shown in Fig.$~$\ref{Fig:alfpminalfinvall}. Such a correlation,
which may look straightforward, has not been explored much 
in the past. We note that this correlation is present for all but a few
objects located at low values of  $\alpha_{\rm inv}$ and $\alpha(P_{\rm min})$
(one of them being (44) Nysa), and
that the location of the Barbarians corresponds to the extrapolation
of the linear trend exhibited by ``normal'' asteroids.
 
The interpretation of the features discussed so far is not immediately obvious, but 
we believe this could be a useful input for current models of light scattering
from rocky and/or icy planetary surfaces.

\begin{figure}
\begin{center}
\includegraphics[width=88mm]{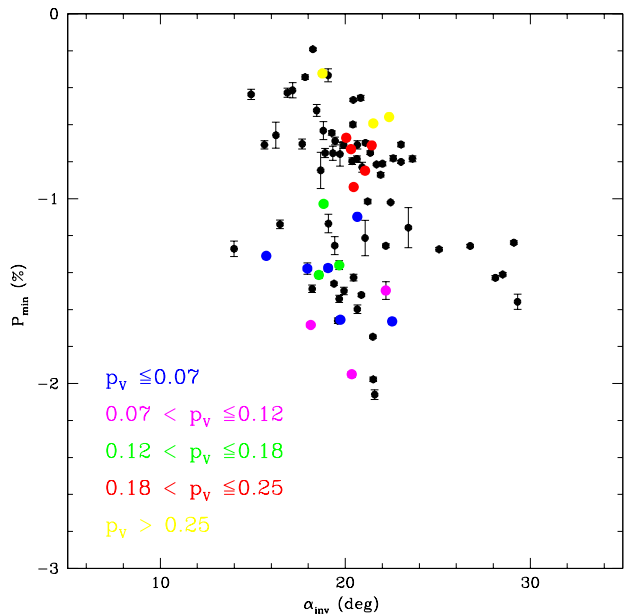}
\end{center}
 \caption{$P_{\rm min}$ \emph{versus} inversion angle $\alpha_{inv}$ for the whole 
sample of asteroids listed in this paper and in Paper I. Objects having albedo values 
determined by \citet{ShevTed} are plotted in color. Different colors correspond to 
different intervals of albedo, as specified in the Figure.}
 \label{Fig:pmininv}
\end{figure}
Finally, we show in Fig.$~$\ref{Fig:pmininv} the relation between $\alpha_{\rm inv}$ and 
$P_{\rm  min}$. This is a classical relation analyzed in the past by different authors
to derive information on the likely properties of the particles which
typically form the surface regolith of the asteroids \citep[for a
  classical review, see][]{Dollfus89}. Again, our analysis  
includes not only the objects listed in Table 1, but also the
$22$ asteroids included in the \citet{ShevTed} list that we
considered in Paper I. In  Fig.$~$\ref{Fig:pmininv} the objects for which 
we have an albedo value given by
\citet{ShevTed} are indicated by color symbols, using different colors
for different albedo classes.

Figure \ref{Fig:pmininv} can be considered as an
updated version of the analogous Figure shown by \citet{Dollfus89} in
the \emph{Asteroids II} book. We note that \citet{Dollfus89} showed
that asteroid data in the $P_{\rm min}$ - $\alpha_{inv}$ plane are
found in a domain which is intermediate between one occupied by
coarse rocks and one occupied by very thin lunar fines composed of
particles smaller than $30$ $\mu$m, according to laboratory
measurements.  We confirm that most asteroids of our sample occupy the
region already found by \citet{Dollfus89}. Asteroids of increasingly
higher albedo tend to occupy regions at the top of the asteroid
domain, but there is some mixing at low albedo values, with some
objects having small \citet{ShevTed} albedos, below $0.07$, which are
found in this plot mixed with asteroids having albedos larger than
$0.12$. One of the mixed objects is (2) Pallas, with its
albedo of $0.145$ according to \citet{ShevTed}, an unexpected value
for an object belonging to the $B$-class, as discussed in Paper I. 

The major difference with respect to the classical results by
\citet{Dollfus89}, however, is the presence of some objects which are
located well outside the typical domain of asteroids, which are found
much closer to or within the domain found by \citet{Dollfus89} for very
pulverized material. These asteroids are Barbarians: (234) Barbara,
(172) Baucis, (387) Aquitania and (980) Anacostia, which 
all have inversion angles above $28^\circ$, and (236) Honoria, with an
inversion angle above $26^\circ$. These objects occupy clearly
anomalous locations in Fig.$\:$\ref{Fig:pmininv}.  Another object with
a relatively high value of the inversion angle ($25^\circ$) in
Fig.$\:$\ref{Fig:pmininv} is (21) Lutetia. With a $P_{\rm min}$ value
of about $-1.27$, this asteroid would also be located in the domain of
very pulverized rocks and lunar fines, according to \citet{Dollfus89}.
Observations carried out from the ground
and by the Rosetta probe during its fly-by of Lutetia have already
provided evidence that this asteroid is unusual in several
respects. It is encouraging, however, to mention that, according to
\citet{Keihmetal2012}, the thermal inertia of Lutetia is
quite low, in very good agreement with the hypothesis that its surface
could be rich in fine dust. Any further attempt of interpretation,
however, must be postponed to future investigations.
%
\section{Conclusions and future work}\label{conc}
The results of an extensive analysis of available asteroid
polarimetric data, carried out in this paper and in Paper I, 
confirm that the study of the polarimetric properties of
these objects is extremely interesting and a powerful
tool for their physical characterization. 

Several results shown in the previous Sections 
deserve further studies mainly on the theoretical side,
because we now have a wealth of information to test and extend current 
models of light-scattering phenomena. In particular, we find that some
features of the negative polarization branch of phase - polarization
curves (distributions of $\alpha_{\rm inv}$ and $\alpha(P_{\rm min})$ 
and the mutual relation between these parameters) are particularly 
interesting. The location of Barbarian asteroids in the $P_{\rm min}$ -
$\alpha_{\rm min}$ plane suggests that their surfaces are covered by 
extremely fine dust particles. 

In this paper and Paper I we give
albedo values obtained from polarimetric parameters for a data-set of
$86$ asteroids. This data-set will hopefully increase rapidly in the years to
come, as an effect of new campaigns of polarimetric observations.
Some problems are certainly still open, including the apparently 
very high albedo of asteroid (44) Nysa. This is not atypical, as in general, 
high-albedo, $E$-class asteroids tend to display a rather large variation 
of albedo, depending on the choice of the polarimetric parameter adopted 
to obtain it. This implies that
the calibration of the relation between geometric albedo and 
polarization parameters could yet see some further improvement.

We expect progress in the field to come in the near future 
from different directions. One possible development would be a systematic 
use of spectro-polarimetry. This will allow observers to profit from 
the results from spectroscopy and polarimetry, separately, plus
the product of this merging of two separate techniques, namely the
study of the variation of the linear polarization as a function of
wavelength. Pioneering work in this respect has been already done
by \citet{Irina09} using broad-band polarimetric data obtained in
different colours, while some of us have recently started a systematic
campaign of spectro-polarimetric observations of asteroids at the VLT
and WHT telescopes which has already provided encouraging
results \citep{Bagnulo14}.

Finally, another essential input is provided by the 
\emph{in situ} observations of the asteroid (4) Vesta performed by the
Dawn spacecraft. Vesta is the only asteroid for which a clear variation
of the degree of linear polarization as a function of rotation has
been convincingly demonstrated 
\citep[see][and references therein]{Dollfus89}. 
Some of us have recently carried out an extensive analysis of available 
``polarimetric lightcurve'' data of (4) Vesta, by
computing the location of the sub-Earth point on
Vesta at the epoch of different, ground-based polarimetric observations, 
in order to understand the relations with the
varying, average properties of the surface seen by ground-based observers
at different epochs, and to look for possible correlations with the
albedo, topography and composition \citep{Vesta}. 
In this study we provided the first example of ``ground-truth'' in asteroid 
polarimetry.

\section*{Acknowledgements} 
AC was partly supported by funds of the PRIN INAF 2011. AC and SB acknowledge support from COST Action MP1104  
``Polarimetry as a tool to study the solar system and beyond'' through funding granted for Short Terms Scientific 
Missions and participations to meetings. RGH gratefully acknowledges financial support by CONICET through PIP
114-200801-00205. We thank the Referee, V. Rosenbush, for her careful review.

\label{lastpage}
\end{document}